\documentclass[aps,prb,nofootinbib,floatfix,twocolumn,showpacs]{revtex4}

\usepackage{times}
\usepackage{color}
\usepackage{graphicx}
\usepackage{dcolumn}
\usepackage{amssymb}
\usepackage{amsmath}
\usepackage{amsfonts}
\usepackage[colorlinks=true,linkcolor=blue,pagecolor=blue,filecolor=blue,menucolor=blue,urlcolor=blue,citecolor=blue,anchorcolor=blue]{hyperref}%

\newcommand{\ie}{\textit{i.e. }}
\newcommand{\eg}{\textit{e.g. }}
\newcommand{\etal}{\emph{et al.}}
\newcommand{\sns}{S/N/S }

\begin{document}

\title{$\varphi$-State and Inverted Fraunhofer Pattern
in Nonaligned Josephson Junctions}

\author{Mohammad Alidoust }
\email{phymalidoust@gmail.com} \affiliation{Department of Physics,
Norwegian University of Science and Technology, N-7491 Trondheim,
Norway}
\author{Jacob Linder}
\email{jacob.linder@ntnu.no} \affiliation{Department of Physics,
Norwegian University of Science and Technology, N-7491 Trondheim,
Norway}
\date{\today}

\begin{abstract}
A generic nonaligned Josephson junction in the presence of an
external magnetic field is theoretically considered and an unusual
flux-dependent current-phase relation (CPR) is revealed. We explain
the origin of the anomalous CPR via the current density flow induced by the external field within a two-dimensional quasiclassical
Keldysh-Usadel framework. In particular, it is demonstrated that
nonaligned Josephson junctions can be utilized to obtain a
ground-state other than 0 and $\pi$, corresponding to a so-called
$\varphi$-junction, which is tunable via the external magnetic flux.
Furthermore, we show that the standard Fraunhofer central peak of
the critical supercurrent may be inverted into a local minimum
solely due to geometrical factors in planar junctions. This yields
good consistency with a recent experimental measurement displaying
such type of puzzling feature [R. S. Keizer \etal,
\href{http://www.nature.com/nature/journal/v439/n7078/full/nature04499.html}{\newblock
Nature \textbf{439}, 825 (2006)}].

\end{abstract}

\pacs{74.50.+r, 74.45.+c, 74.25.Ha, 74.78.Na }

\maketitle

A sinusoidal current-phase relation (CPR) and Fraunhofer response of
the critical supercurrent through a $s$-wave Josephson contact
exposed to an external magnetic field are often considered to be
standard characteristics of such junctions
\cite{cite:josephson,cite:bergeret,cite:clarke,cite:Barzykin}.
Nevertheless, several theoretical studies have been dedicated to the
aim of achieving an experimentally accessible situation where the
CPR is non-sinusoidal
\cite{cite:buzdin1,cite:buzdin2,cite:goldobin}. In this case, the
Josephson ground-state may be characterized by an arbitrary
superconducting phase difference $\varphi$
\cite{cite:buzdin1,cite:buzdin2,cite:goldobin,cite:barash,cite:bakurskiy},
rather than the so-called $0$- and $\pi$-states \cite{cite:bula}.
The first experimental realization of such a $\varphi$-junction was
very recently reported in Ref. \onlinecite{cite:sickinger}.

\begin{figure}[t!]
\includegraphics[width=8.50cm,height=5.0cm]{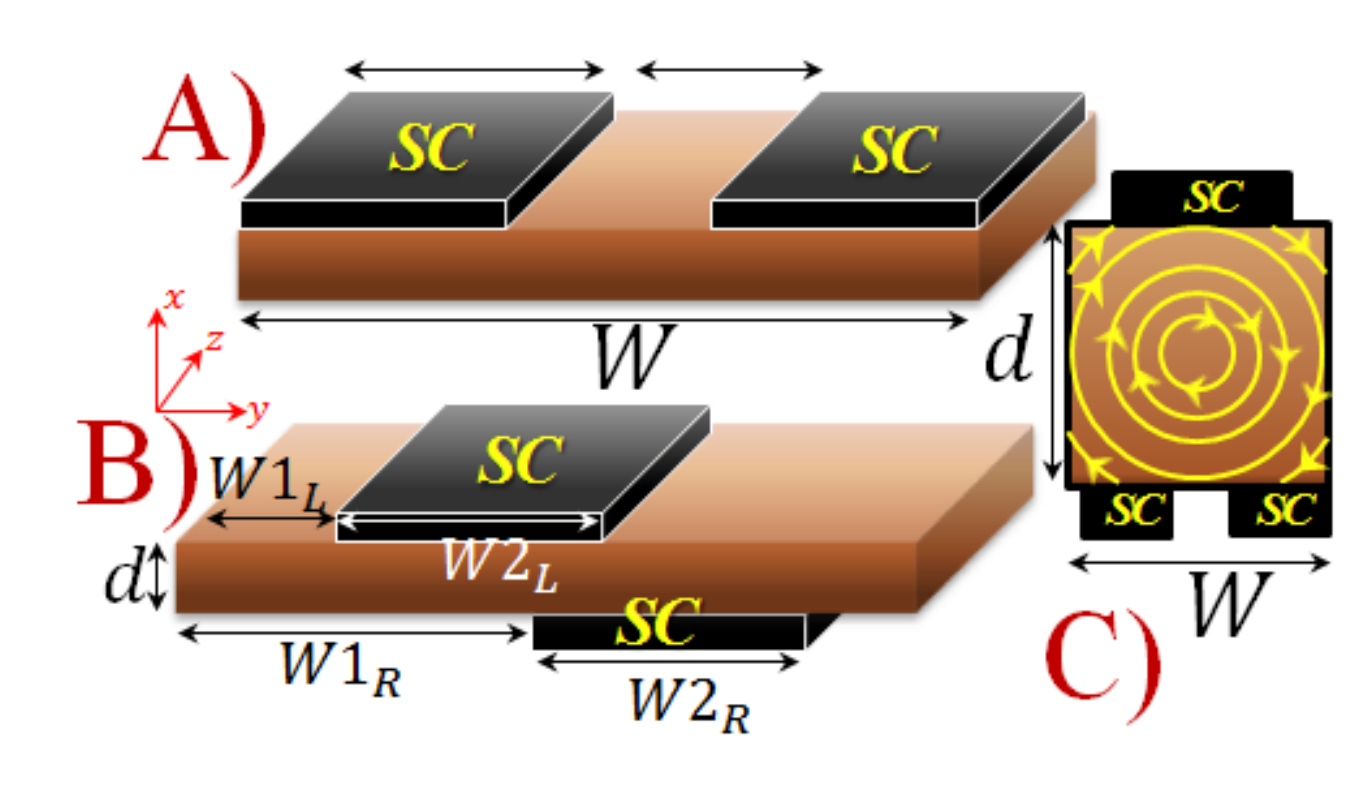}
\caption{\label{fig:model} Diagram of considered setups in this
paper. An external magnetic field \textbf{H} (not shown) is applied
to the junction in the $z$ direction. The junction lengths and widths are $d$ and $W$,
respectively. A): The planar Josephson junction that has
experimentally been studied in \eg Ref. \onlinecite{cite:keizer}.
The widths of the superconducting leads are assumed to be $W1_L$ and
$W-(W1_L+W2_L)$. B): The usual stacked geometry of a Josephson
junction with displaced superconducting leads. The superconducting
leads' sizes are $W1_L$ and $W2_R$ at the top and bottom of
junction, respectively. C): Qualitative view of the current density flow inside the normal strip subject to an external
magnetic field, which is used to describe the origin of the
addressed unusual CPR.}
\end{figure}

Recent studies have also pointed to the fact that the conventional
Fraunhofer pattern in Josephson junctions may be modified by the
junction geometry or interfacial pair breaking
\cite{cite:alidoust,cite:bergeret,cite:barash}. The suppression of
the central peak in the interference pattern can also occur in
systems consisting of a superposition of multiple $0$-$\pi$
junctions \cite{cite:alidoust,cite:goldobin,cite:barash}. However,
there still exists experimentally observed magnetic interference
profiles that remain unsettled in terms of a theoretical explanation
of the physical origin \cite{cite:keizer,cite:norman}. In
particular, Keizer \etal~ \cite{cite:keizer} observed an anomalous
interference pattern with a local minimum at zero flux in addition
to slowly damped oscillations of the critical supercurrent compared
to the standard Fraunhofer pattern. The setup in Ref.
\onlinecite{cite:keizer} consisted of a planar Josephson junction
where superconducting leads were deposited on a same side of a
half-metallic ferromagnetic strip which was fully spin polarized.
Figure \ref{fig:model} A) depicts diagrammatically the mentioned
experimental setup. To study the system theoretically, Ref.
\onlinecite{cite:mohammadkhani} utilized the Eilenberger formalism
in a ballistic planar junction, similar to the setup of Ref.
\onlinecite{cite:keizer}, while neglecting the orbital motion \cite{cite:zagoskin} of the quasiparticles.
Consequently, an effective spatially dependent superconducting phase
difference was obtained via Ginzburg-Landau theory and substituted
into the Eilenberger equation. An almost $\Phi_0$-periodic pattern
with non-zero minima of the critical current with respect to
external magnetic flux were found due to the appearance of second
harmonic ($\sin2\varphi$, see also Ref. \onlinecite{cite:buzdin1}).
However, the inverted interference pattern with a local minimum at
zero flux was not reproduced.

\begin{figure}[t!]
\includegraphics[width=8.5cm,height=5.5cm]{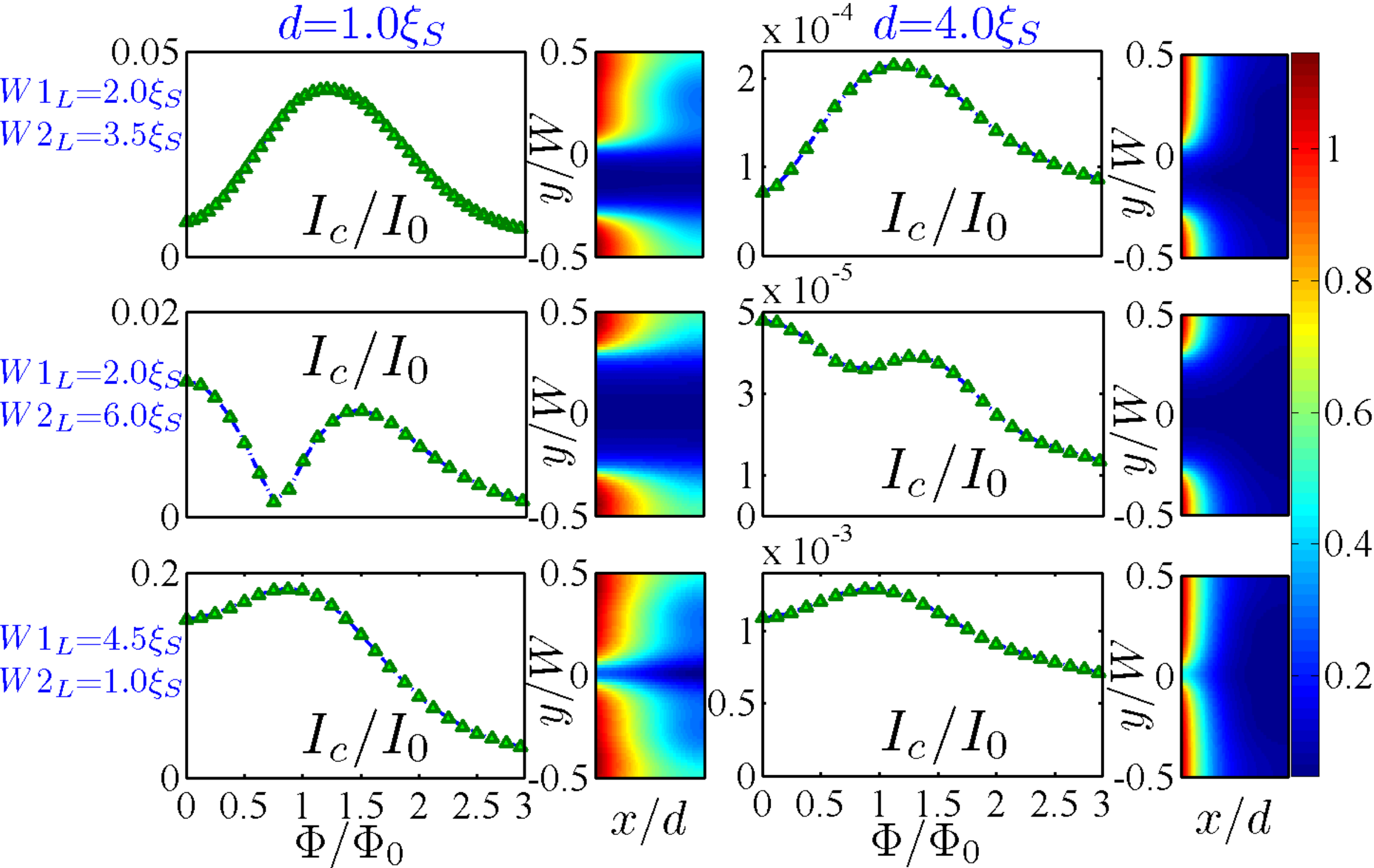}
\caption{\label{fig:Wl} Critical supercurrent as a function of
external magnetic flux $\Phi$ through the normal part of junction.
The corresponding pair potential spatial map is given with
$\varphi=0$. Throughout the paper we have assumed that the junction
width is fixed at $W$=$10\xi_S$. The first and second columns show
the critical current $I_c/I_0$ vs normalized external magnetic flux
$\Phi/\Phi_0$ and the corresponding pair potential spatial maps with
thicknesses $d$=$\xi_S$ and $4\xi_S$, respectively.
Each row indicates different values of $W1_L$ and $W2_L$ namely, the
first superconducting lead size and the separation of the
superconducting leads, respectively (see Fig. \ref{fig:model}).}
\end{figure}

In this Rapid Communication, we consider a generic class of Josephson junctions in
the presence of an external magnetic field where the position of the
superconducting leads relative each other is not necessarily aligned
(see Fig. \ref{fig:model}). The obtained results are derived
\textit{without recourse to any ansatz} - we have instead utilized a
quasiclassical Keldysh-Usadel technique with the numerical approach
in Ref. \onlinecite{cite:alidoust} and solved exactly the resultant
linearized equations of motion for the Green's function. As our
\textit{first} main result, we unveil that the origin of the
unexpected interference pattern in the experiment of Ref.
\onlinecite{cite:keizer} lies within the geometry of the setup. In
this way, the absence of the standard Fraunhofer pattern, which has
not been clearly understood, is resolved. In addition to this, we
demonstrate as our \textit{second} main result that the CPR in
nonaligned junctions takes on a very unusual feature: it becomes
shifted by a term proportional to the external flux $\Phi$, namely
$I(\varphi,\Phi) = I_0(\Phi)\sin(\varphi+\Theta(\Phi))$ where
$\varphi$ is the superconducting phase difference and $\Theta$ is a
geometry-dependent function. Our investigations reveal that the
well-known sinusoidal supercurrent and consequently the Fraunhofer
pattern manifest \textit{only in a specific} situations. This result
is explained in terms of the current density flow stemming
from the orbital effect induced by the magnetic field. An interesting consequence of the external
magnetic flux-shifted superconducting phase-difference is that the
ground-state of the system may be tuned via the external field so
that the equilibrium phase difference differs from the conventional
$0$ or $\pi$ solutions making a so-called $\varphi$-junction. This
might constitute a simpler alternative to realizing a
$\varphi$-state compared to the array of SFS junctions considered in
Ref. \onlinecite{cite:sickinger}.

\begin{figure*}[t!]
\includegraphics[width=17cm,height=5.7cm]{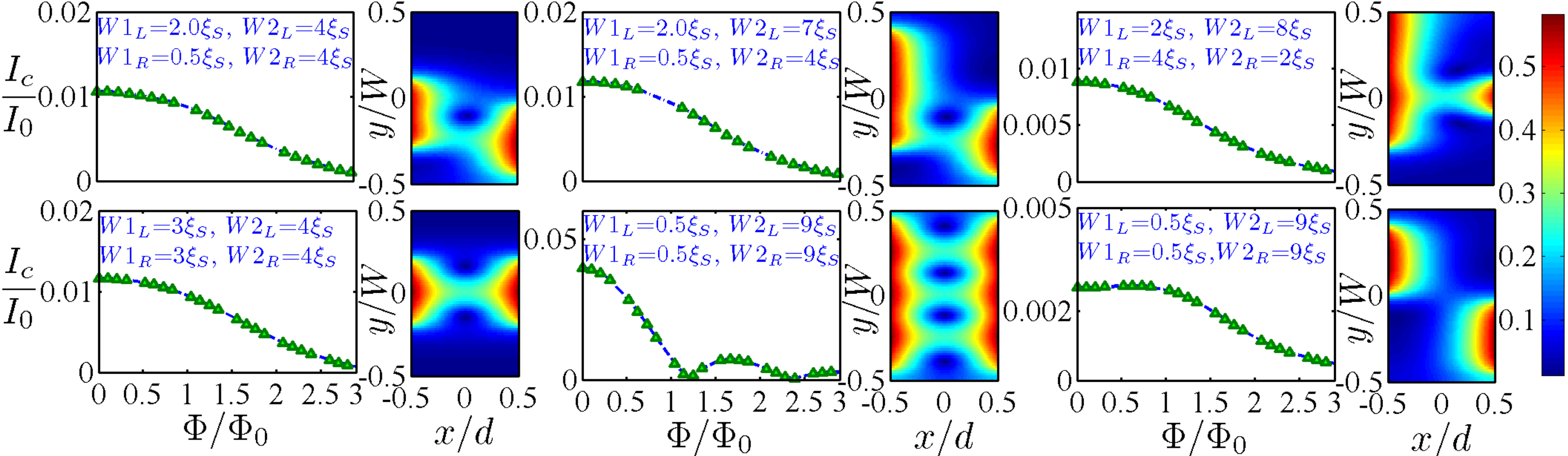}
\caption{\label{fig:WLWR} Critical supercurrent against external
magnetic flux and corresponding pair potential spatial maps of standard (stacked)
Josephson junctions with displaced superconducting leads including
various lead sizes. For the pair potential maps, the superconducting
phase difference and external magnetic flux are fixed at $\varphi$=0
and $\Phi$=$4\Phi_0$, respectively. The junction thickness and width
are set to $d$=$2\xi_S$ and $W$=$10\xi_S$, respectively.}
\end{figure*}

In the presence of impurity scattering, i.e. the diffusive regime of transport, the quasiparticles' momentum is integrated
over all directions in the space which leads the Usadel equation.
Solving the Usadel equation in the presence of a magnetic field allows one to compute the current density flow profile in the junction which is different from the
individual trajectoriers taken by each quasiparticle.
Grazing trajectories are not well-defined in this regime although
they need to be considered carefully in the clean regime
(where the Eilenberger equation is valid) \cite{cite:zagoskin}.

The
starting point for the analysis is the equation of motion for the
Green's function in the diffusive regime provided by the Usadel
equation \cite{cite:usadel}:
\begin{eqnarray}
D[\hat{\partial},\check{G}(x,y,z)[\hat{\partial},\check{G}(x,y,z)]]+i[
\varepsilon \hat{\rho}_{3} ,\check{G}(x,y,z)]=0,
\end{eqnarray}
where $\hat{\rho}_{3}$ is $4\times 4$ Pauli matrix. Here,
$\varepsilon$ is the particles' energy measured from the Fermi level
and $D$ is medium diffusive constant. In the presence of an external
magnetic field $\textbf{H}$ and its vector potential $\textbf{A}$,
$\hat{\partial}\equiv\vec{\nabla} \hat{1}-ie
\mathbf{A}(x,y,z)\hat{\rho}_{3} $ provided that \cite{cite:morten}
\begin{eqnarray}
[\hat{\partial},\hat{G}(x,y,z)]=\vec{\nabla}\hat{G}(x,y,z)-ie[\mathbf{A}(x,y,z)\hat{\rho}_{3},\hat{G}(x,y,z)].
\end{eqnarray}
The vector potential is an arbitrary quantity except for the
restriction $\vec{\nabla}\times\textbf{A}=\textbf{H}$. We use the
Coulomb gauge $\vec{\nabla}\cdot\mathbf{A}=0$ throughout our
calculations and assume that the external magnetic field is oriented
in the $z$ direction \ie $\textbf{H}=H\hat{z}$ (see Fig.
\ref{fig:model}). Thus, we may use $\textbf{A}=-yH\hat{x}$. In
general, the Usadel equation should be simultaneously solved along
with the Maxwell equation $\vec{\nabla}\times\mathbf{H}=\mu_{0}\mathbf{j}$ in
a self-consistent manner to take into account the influence of
screening currents. The experimentally relevant scenario is
considered where the width of the junction $W$ is smaller than the
Josephson penetration length $\lambda_{J}$, allowing us to ignore
the screening of the magnetic field
\cite{cite:bergeret,cite:alidoust,cite:crosser}. The Usadel motion
equation yields a system of nonlinear coupled complex partial
differential equations that should be supported by suitable boundary
conditions for studying junctions. In our Josephson system, we
employ the Kupriyanov-Lukichev boundary conditions at N/S interfaces
\cite{cite:zaitsev} and control the leakage of superconductive
correlations into the normal strip using an interface parameter
\begin{figure}[t!]
\includegraphics[width=8.5cm,height=8cm]{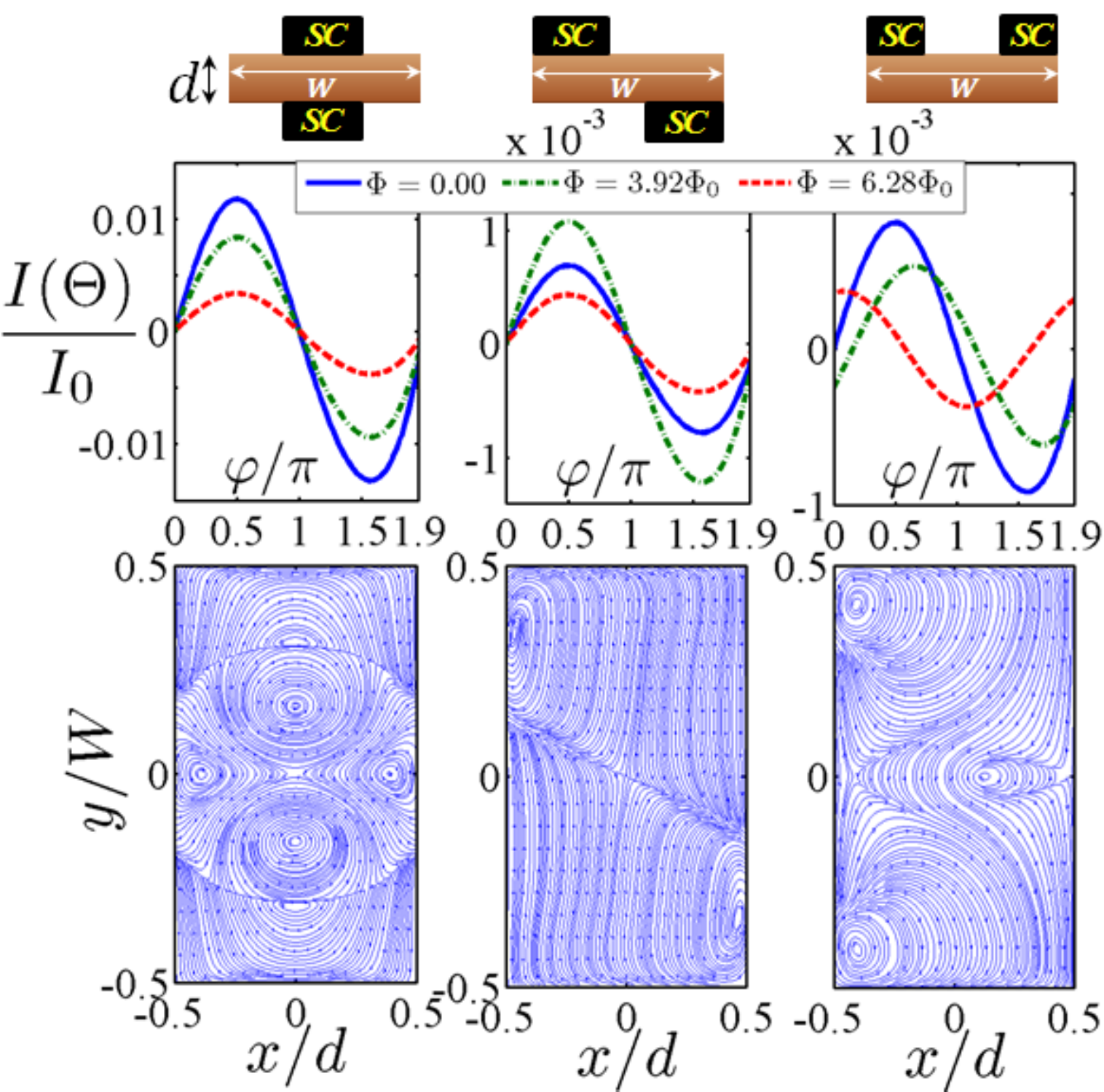}
\caption{\label{fig:crntdnsty} \textit{i)} Left column:
$W1_L=3\xi_S$, $W2_L=4\xi_S$, $W1_R=3\xi_S$, $W2_R=4\xi_S$,
\textit{ii)} middle column: $W1_L=6\xi_S$, $W2_L=4\xi_S$, $W1_R=0$,
$W2_R=4\xi_S$ and finally \textit{iii)} right column: $W1_L=2\xi_S$,
$W2_L=6\xi_S$. The top panels represent the CPRs for various values
of $\Phi/\Phi_0$=$0$, $3.92$, $6.28$. The current density spatial
maps in the bottom row show the results for $\varphi=0$ and
$\Phi=4\Phi_0$. The superconducting leads' sizes are set equal at
$4\xi_S$ for all cases as schematically depicted on top of each
column.}
\end{figure}
$\zeta$;
\begin{eqnarray}
\zeta(\hat{G}(x,y,z)\hat{\partial}\hat{G}(x,y,z))\cdot\check{\boldsymbol{n}}=[\hat{G}_{\text{BCS}}(\varphi),\hat{G}(x,y,z)],
\end{eqnarray}
in which $\hat{\boldsymbol{n}}$ is a unit vector denoting the
perpendicular direction to an interface and $\varphi$ is the bulk
superconducting macroscopic phase. We define $\zeta=R_B/R_F$ as the
ratio between the resistance of the barrier region and the
resistance in the normal sandwiched strip. The bulk solution for the
retarded Green's function in a $s$-wave superconductor is given by
\cite{cite:morten} $g^{R}_{\text{BCS}}=\cosh\vartheta(\varepsilon)$
and $f^{R}_{\text{BCS}}=e^{i\varphi}\sinh\vartheta(\varepsilon)$ in
which $\vartheta(\varepsilon)=\text{arctanh}(|\Delta|/\varepsilon)$.
For a weak proximity effect ($\zeta\gg1$), the normal and anomalous
Green's functions can be approximated by
 $g^{R}\simeq 1$
and $|f^{R}|\ll 1$, respectively. The current density vector is
expressed via the Keldysh block as
\begin{eqnarray}
\mathbf{J}\text{(}\vec{R},\varphi\text{)}=J_{0}\int
d\varepsilon\text{Tr}\{\rho_{3}(\hat{G}(x,y,z)[\hat{\partial},\hat{G}(x,y,z)])^{K}\}.
\end{eqnarray}
Here, $J_{0}$ is a normalization constant proportional to the density of states $N_{0}$ at the
Fermi level. The total supercurrent $I$ is obtained by integrating the current density over the interface area of the superconducting banks. The flux penetrating the junction is given by
$\Phi=dWH$. We also investigate the spatial variation of pair
potential inside the normal region calculated via:
\begin{eqnarray}
U=U_{0}\text{Tr}\{(\hat{\rho}_{1}-i\hat{\rho}_{2})\int d\varepsilon
\hat{\tau}_{3}\check{G}^{K}(x,y,z)\},
\end{eqnarray}
where $U_{0}=-N_{0}\lambda/16$ \cite{cite:morten}. In the
presence of an external magnetic field, the resultant differential
equations and boundary conditions have a more complicated coordinate-dependence which
renders an analytical solution virtually impossible. Without any orbital effect, such a solution may be obtained \cite{cite:bakurskiy}. To study the considered Josephson junction
we use a collocation finite element numerical method the same as
Ref. \onlinecite{cite:alidoust}. The components of approximate
solution are assumed to be linear combinations of bicubic Hermite
basis functions satisfying the boundary conditions. Ultimately, the
resultant nonsymmetric linear algebraic equations are solved via a
Jacobi conjugate-gradient method. For more details, see Ref.
\onlinecite{cite:sewell2a}. All lengths and energies are normalized
by the superconducting coherent length $\xi_S$ and superconducting
gap at absolute zero $\Delta_0$. The barrier resistance $\zeta$ is
fixed at $7$ ensuring the validity of weak proximity regime.
Temperature and junction width are $T=0.05T_c$ and $W=10\xi_S$. We
use units such that $\hbar=k_B=1$.

Figure \ref{fig:Wl} illustrates the response of the critical
Josephson current in a planar junction to an external magnetic field
as shown schematically in Fig. \ref{fig:model} A). Various parameter
values have been considered in order to make our analysis as general
as possible. To do so, we have considered three scenarios where the
superconducting leads have different sizes (first row) and where
they have equal sizes with a large (second row) and small (third
row) separation distance. Specifically, the third row is relevant
with regard to the experiment in Ref. \onlinecite{cite:keizer} where
the size of the electrodes far exceeds the separation distance. As
seen, in this case the interference pattern exhibits a local minimum
at $\Phi=0$ rather than a maximum as in the Fraunhofer case, which
is fully consistent with the experimental results in Ref.
\onlinecite{cite:keizer}. Whereas it was speculated that this
minimum might be attributed to a shift in the entire interference
curve due to a finite sample magnetization in Ref.
\onlinecite{cite:keizer}, it is obvious that this is not the case
here since the sandwiched strip is not ferromagnetic. Moreover, such
a shift would make the current vs. flux curve manifestly asymmetric
(see \eg Ref. \onlinecite{cite:robinson}), in contrast to the
experimental results of Ref. \onlinecite{cite:keizer} where the
central minimum is flanked by two large peaks, similar to our
results. Based on this, it seems reasonable to explain the deviation
from the standard Fraunhofer pattern as a result originating from
the combination of a planar geometry with the size and separation
distance of the superconducting electrodes. The latter fact is seen
by considering the second row of Fig. \ref{fig:Wl} where the
separation distance is large compared to the superconductors: a
Fraunhofer-like pattern emerges, although the decay becomes more
monotonic as the thickness $d$ of the normal strip increases. Even
columns in both Figs. \ref{fig:Wl} and \ref{fig:WLWR} show the pair
potential where the superconducting phase difference is zero
$\varphi=0$ and external magnetic flux is set to $\Phi=4\Phi_0$. As
seen, the predicted proximity vortices in Refs.
\onlinecite{cite:bergeret} and \onlinecite{cite:alidoust} vanish for
the planar junction geometry. However, as it will be discussed
further below, they reappear in the specific case of a stacked
geometry (Fig. \ref{fig:model} B).

It is instructive to contrast these results with the geometry of
Fig. \ref{fig:model} B) where the two superconducting leads are
connected to the normal strip at opposite edges. This is resemblant
to the experimentally often used stacked geometry. The order of
frames (critical current and corresponding pair potential spatial
map) are identical to those in Fig. \ref{fig:WLWR} and various lead
sizes and locations are investigated. It is seen that the location
and size of both terminals are vital in terms of determining how the
critical current responds to the external flux. For instance, our
results reveal that only in specific case where the width of the
leads' are sufficiently large and connected to opposite edges
precisely in front of each other does one recover a
proximity-induced vortex pattern along with the Fraunhofer curve \ie
$I(\varphi,\Phi) \propto \Phi^{-1}\sin\Phi\sin\varphi$ which is a
special case corresponding to the scenario of Ref.
\onlinecite{cite:bergeret}. The results for the other scenarios in
Fig. \ref{fig:WLWR} also show good consistency with previous
experimental observations \cite{cite:angers}.

It is worth examining the characteristic length scales and thus the
radius of the current circulation in Fig. \ref{fig:crntdnsty}. To
illustrate this, we consider for concreteness the simplest case of a
wide \sns junction subject a perpendicular magnetic field (to see
more details, see Ref. \onlinecite{cite:bergeret}). In this
particular case, the current density is given by
$\mathbf{J}_x\text{(}\vec{R},\varphi\text{)}=J_{0x}\sin(\varphi-2\frac{\pi\Phi
}{\Phi_0W}y)$. As seen, the characteristic length scale $L_c$ over
which the current density changes upon moving along the $y$ axis is
$L_c \sim \Phi_0W/\Phi$. Thus, for magnetic fields corresponding to
several flux quanta $L_c$ can be smaller than the junction size.
With increasing external magnetic flux $\Phi$, the current density
flow shown in Fig. \ref{fig:crntdnsty} takes on smaller radii.
Instead, when decreasing the external flux $\Phi\rightarrow 0$,
$L_c\rightarrow\infty$ which means there exists no current
circulation in the system. In other words, the current density
spatial map of the system is uniform in the absence of any external
magnetic flux.

Having unveiled the origin of the anomalous inverted Fraunhofer
response, we now turn to the second main result of this paper: the
possibility to generate a $\varphi$-junction in an SNS system with
an applied magnetic field. In Fig. \ref{fig:crntdnsty}, we provide
the CPR in addition to a spatial map of the current-flow in the
normal strip for three represented geometries. In \textit{i)} the
leads are connected opposite to each other, in \textit{ii)} they are
connected antisymmetrically, whereas in \textit{iii)} they are
connected symmetrically in a planar geometry similar to Ref.
\onlinecite{cite:keizer}. It is clear that the CPR remains
sinusoidal as a function of the superconducting phase-difference
$\varphi$ in both \textit{i)} and \textit{ii)} independent on the
applied flux. However, case \textit{iii)} is qualitatively
different. The generic form of the CPR is now revealed as:
\begin{align}\label{eq:cpr}
I(\varphi,\Phi)=I_0(\Phi)\sin(\varphi+\Theta(\Phi))
\end{align}
in which $I_0(\Phi)$ and $\Theta(\Phi)$ are geometry-dependent
functions of external magnetic flux as seen in Fig.
\ref{fig:crntdnsty}. In fact, Eq. (\ref{eq:cpr}) holds for all
situations considered in Fig. \ref{fig:Wl} where we have
demonstrated the CPR is never purely sinusoidal. The standard
sinusoidal CPR is recovered only for symmetric situations relative
the induced orbital motion by the external magnetic field, see Fig.
\ref{fig:model} C). This observation has a highly interesting
consequence: the anomalous magnetic flux-coupled CPR ensures that
the ground-state of the system may be tuned so that the equilibrium
phase difference differs from the conventional 0 or $\pi$ solutions.
Instead, a so-called $\varphi$-junction may be realized where the
ground-state phase difference $\varphi$ is tunable via the external
flux. We therefor arrive at a ground-state with Josephson energy
$E_J$ which can be controlled
by adjusting the applied external magnetic field. The idea of a
$\varphi$-junction via a superconducting phase difference shift has
been considered previously \cite{cite:buzdin2} in the context of a
non-centrosymmetric normal layer with a Rashba spin-orbit
interaction. However, in our setup the external flux is a
well-controlled parameter which allows for easy tuning of the
ground-state, as opposed to controlling a spin-orbit interaction
parameter. Moreover, our finding is different from Ref.
\onlinecite{cite:goldobin} where two magnetic junctions, one in
$0$-state and the other in $\pi$-state with different lengths, are
connected in parallel and consequently generate an extra
cosinusoidal term in addition to negative second harmonic.

What is then the physical origin of this anomalous CPR? The answer
to this question may be obtained by investigating the current density flow under the influence of an external magnetic
field inside the normal strip, as seen in Fig. \ref{fig:crntdnsty}. For zero phase difference $\varphi=0$, the
external magnetic field induces a current flow where the orbital
paths taken by the quasiparticles move with the same flux in and out
of the superconducting regions, in effect no net current flow,
\textit{only in special geometrical configurations}. For instance,
both in \textit{i)} and \textit{ii)} the current flow between the
superconductors in any part of the normal region is seen to have an
antisymmetric, and thus cancelling, contribution in a different part
of the normal strip at zero phase difference $\varphi=0$. In
contrast, this is no longer the case in setup \textit{iii)}: there
is a net flow of current induced by the orbital response due to the
magnetic flux, even at $\varphi=0$. To elucidate this clearly in the
current-flow, one would have to consider the amplitude of the local
current as well, but the supercurrent-phase curves nevertheless
demonstrate that this interpretation is correct. In essence, this is
a geometry-dependent effect since it relies on the positioning of
the leads relative the induced current-flow via the applied field.
Thus, it gives rise to the unique possibility to alter the standard
CPR so that the ground-state of system can be adjusted by tuning the
external flux.


To conclude, we have studied the Josephson critical current and its
response to an external magnetic flux in experimentally feasible
nonaligned junctions. Specifically, a planar geometry similar to a
recent experiment \cite{cite:keizer} is considered and it is
demonstrated that the observed suppression at zero flux may stem
from the junction geometry rather than any intrinsic magnetization.
Moreover, it is shown that a highly unusual supercurrent-phase
difference-shift occurs inevitably in a class of nonaligned
junctions due to an external magnetic flux. Its precise form is
sensitive to the size and location of the superconducting leads.
Consequently, this offers a route to a tunable junction
ground-state. The physical origin of this effect is traced back to
the induced current density flow due to the presence of an
external field relative the position of the superconducting leads.
As an interesting consequence, this type of Josephson junctions
constitute an attainable way of realizing the so-called
$\varphi$-junction experimentally.

We appreciate G. Sewell for his valuable instructions in the
numerical parts of this work. We also thank F. S. Bergeret, E.
Goldobin, J. W. A. Robinson, V.V. Ryazanov and N. Birge for useful
discussions/comments as well as K. Halterman for his generosity
regarding compiler source.

\end{document}